# Efficient *n*-type Doping in Epitaxial Graphene through Strong Lateral Orbital Hybridization of Ti Adsorbate

Jhih-Wei Chen[1], Hao-Chun Huang[1], Domenica Convertino[2], Camilla Coletti[2], Lo-Yueh Chang[3,4], Hung-Wei Shiu[4], Cheng-Maw Cheng[4], Ming-Fa Lin[1], Stefan Heun[5], Forest Shih-Sen Chien[6], Yi-Chun Chen[1], Chia-Hao Chen[4*], and Chung-Lin Wu[1*]

[1]Department of Physics, National Cheng Kung University, Tainan 70101, Taiwan

[2]Center for Nanotechnology Innovation @ NEST, Istituto Italiano di Tecnologia, Pisa 56127, Italy

[3]Department of Physics, National Tsing Hua University, Hsin Chu 30013, Taiwan

[4]National Synchrotron Radiation Research Center (NSRRC), Hsinchu 30076, Taiwan

[5]NEST, Istituto Nanoscienze-CNR and Scuola Normale Superiore, 56127 Pisa, Italy

[6]Department of Physics, Tunghai University, Taichung 407, Taiwan

* Correspondence and requests for materials should be addressed to

C. -H. Chen +886-3-5780281-7325. E-mail: chchen@nsrrc.org.tw

C. -L. Wu +886-6-2757575-65219. E-mail: clwuphys@mail.ncku.edu.tw

**Abstract**

In recent years, various doping methods for epitaxial graphene have been demonstrated through atom substitution and adsorption. Here we observe by angle-resolved photoemission spectroscopy (ARPES) a coupling-induced Dirac cone renormalization when depositing small amounts of Ti onto epitaxial graphene on SiC. We obtain a remarkably high doping efficiency and a readily tunable carrier velocity simply by changing the amount of deposited Ti. First-principles theoretical calculations show that a strong lateral (non-vertical) orbital coupling leads to an efficient doping of graphene by hybridizing the $2p_z$ orbital of graphene and the $3d$ orbitals of the Ti adsorbate, which attached on graphene without creating any trap/scattering states. This Ti-induced hybridization is adsorbate-specific and has major consequences for efficient doping as well as applications towards adsorbate-induced modification of carrier transport in graphene.

# 1. Introduction

Achieving new atomic orbital configurations in solid surfaces can lead to novel electronic phenomena, especially in single-layer atom material, such as graphene or other two-dimensional (2D) materials. Examples of properties of graphene that depend on the orbital configuration include surface states [1,2], tunable pseudogap phases [3], and control of carrier doping [4,5,6]. For a honeycomb carbon network, the conduction properties of graphene are determined by the π and π* bands formed by the hybridization of parallel $p_z$ orbitals. When graphene is adsorbing or incorporating selected atoms, the $p_z$ orbitals strongly interact (hybridize) with the atomic orbitals of the latter, and the electronic and transport properties are significantly affected [7,8]. Therefore, the deformed $p_z$ will play a critical role in the diversified electronic properties and might lead to novel properties which are not available in pristine graphene.

In 2D materials, doping achieved through methods such as atom substitution or surface adsorption does not only introduce carriers but also unintentionally affects the physical, chemical, and materials properties. Substitutional doping with different atoms (e.g. B or N) of graphene leads to the distortion of the in-plane $sp^2$ hybridization, and local changes in its chemical bonding and electronic properties [9,10,11,12]. It also introduces scattering centers and/or trapped states which affect the transport properties of graphene [13,14,15]. Recently, adsorption doping was used to alter the electronic structure via appropriate metal atoms [16,17,18]. From a structural viewpoint, to achieve an efficient doping by orbital hybridization between the adsorbed atoms and the C atoms, an adsorbed atom that possesses a shape-consistent orbital to $2p_z$ is highly recommended. In this light, we envision that - as theoretically predicted - a transition metal (TM) atom adsorbed on a $sp^2$-graphene hollow site shall show the lowest adsorption energy, in which the TM 3$d$ orbital and C 2$p_z$ orbital will align vertically. This shall maximize the electronic coupling and, hence, serve as an excellent adsorption doping source because of the good lateral orbital-coupling nature.

In this study we show that by depositing a low amount of Ti atoms on epitaxial graphene on SiC, the electronic band structure of graphene can be significantly engineered (see Fig. 1). Specifically, it is well known that Ti on graphene has a large cohesive energy [19], so that Ti atoms cluster and aggregate rather than forming isolated adatoms [20]. We have therefore explored the Ti low coverage limit in order to obtain well-isolated adatoms on graphene as sketched in panel (a). In this system, a strong orbital hybridization occurs primarily between the vertical Ti-3$d$ and C-2$p_z$ orbitals as a result of symmetry matching and spatial overlap of the orbitals distribution (sketched in panel (b)). We show that the adsorption of Ti atoms onto graphene results in a band structure renormalization, velocity modification, and substantial doping effect as sketched in panel (c). Indeed, angle-resolved photoemission spectroscopy (ARPES) reveals that in the low-coverage regime, Ti adatoms efficiently donate electrons to graphene without causing Fermi level shifting.

## 2. Experimental

ARPES is a powerful tool to *in situ* probe the electronic structure of Ti-adsorbed graphene, which is not available in the transport measurement. Here, a quartz crystal microbalance (QCM) is used to determine the amount of Ti coverage, since QCM has a narrow resonance that leads to highly stable oscillations and a high accuracy, even in the case of extremely small coverages. In order to duplicate the deposition conditions, the QCM was placed in front of the Ti evaporator at the same distance, deposition angle, and base pressure as the sample for the *in situ* ARPES measurements. A commercial "SPECS Phoibos 150 analyzer" is utilized to measure the ARPES spectra. Measurements were performed at the "BL 24 A1" of the National Synchrotron Radiation Research center (NSRRC) in Hsinchu, Taiwan at photon energies of 52 eV. All measurements were carried out at room temperature. Energy and momentum resolution are estimated to be 150 meV and ± 0.005 Å$^{-1}$. Graphene was grown via thermal decomposition in a resistively heated cold-wall reactor (Aixtron HT-BM) on a hydrogen etched 4H-SiC(0001) substrate. The growth process was performed in Ar atmosphere, at 1600 K and 780 mbar for 10 minutes. Number of layers and thickness homogenity were assessed by a combined analysis of Raman spectroscopy and atomic force microscopy [21]. Prior to ARPES measurement, the graphene samples were annealed to 1000 K for a few hours under ultra-high vacuum (UHV) conditions.

**3. Results and discussion**

Fig. 2(a) shows the electronic band structure of epitaxial graphene measured via ARPES upon increasing Ti coverage. The electron pocket that appears near the Fermi level ($E_F$) at the K point is ascribed to the $2p_z$ $\pi^*$ band. As expected for pristine epitaxial graphene, the entire Dirac cone is shifted ~0.4 eV below the Fermi level $E_F$ due to electron doping from SiC substrate [22]. This shows that single-layer pristine epitaxial graphene is used here. We observe that, upon Ti deposition, the Fermi level does not shift: its Dirac point is constant at -360 mV for pristine graphene (0 s) and after 30 s and 60 s of Ti deposition, and -350 mV for 90 s. On the other hand, the energy dispersion deviates in the linear region at lower energies, while the electron pocket is enlarged near the Fermi level. This band renormalization changes with Ti coverage.

We estimate the Ti coverage, and thus correlate band structure changes to the amount of deposited atoms, by means of QCM. Fig. 2(c) shows the Ti thickness calibration for increasing deposition time, from 0 mins to 300 mins. The measured Ti thickness increases linearly with Ti deposition time. By interpolation we estimate the Ti coverage for deposition times of 30, 60s, and 90s to be 1/700, 1/350, and 1/235 ML, respectively.

To better highlight the band renormalization observed upon Ti deposition, we plot in Fig. 2(a) the fits of the momentum distribution curves (MDCs) as a function of energy. At the Fermi level, the intercept of the linear fit yields the value of the Fermi momentum $k_F$, which represents the extent of the K-centered Fermi surface electron pocket along K→Γ. As shown in Fig. 2(a), the value of $k_F$ increases with increasing Ti coverage, from 0.0332 Å$^{-1}$ for pristine graphene to 0.0434 Å$^{-1}$ for a Ti coverage of 1/235 ML. Hence, the charge carrier density – calculated as $n = (k_F)^2/\pi$ – increases from $3.5 \times 10^{12}$ cm$^{-2}$ to $6.0 \times 10^{12}$ cm$^{-2}$ for Ti coverages from 0 ML to 1/235 ML. In addition to the renormalization observed in the electronic dispersion, the line width of the bands increases as well, and this will be discussed quantitatively in Fig. 4.

To quantitatively understand this band renormalization, the Dirac velocity $V_D$ in the low-lying bands is determined. Fig. 2(b) shows that $V_D$ strongly depends on the Ti coverage. As the coverage of Ti increases, the Dirac velocity $V_D$ in the valence band decreases from $1.63 \times 10^6$ m/s to $1.25 \times 10^6$ m/s. In addition to the valence band, the Fermi velocity $V_F$ in the conduction band decreases from $1.65 \times 10^6$ to $1.26 \times 10^6$ m/s as the coverage increases (not shown here). This is consistent with the results presented in Fig. 2(a): due to the linear dispersion relation $|E_F - E_D| = \hbar k_F V_F$, a constant value of $E_F - E_D$ (as observed in Fig. 2(a)) requires that the product of Fermi velocity $V_F$ and Fermi momentum $k_F$ remains approximately constant. Here we observe a decreasing Fermi velocity $V_F$ and an increasing Fermi momentum $k_F$, while their product is approximately constant (to within 2.5%) for the range of Ti coverages explored. We suggest that this approximately constant position of the Dirac point is associated with the small amount of Ti atoms deposited. For larger Ti coverages, McCreary *et al*. observed a shift of the Fermi level [23]. Interpolating their results, the highest Ti coverage used here would result in a shift of the Fermi level of around 140 meV, well below our energy resolution.

Band velocities are slightly asymmetric (i.e. different) between the conduction and valence bands due to the lattice mismatch between epitaxial graphene and the SiC substrate [24,25]. The band velocities modulation reaches ~23%, which is close to the value observed for potassium (K) doped graphene (i.e., 25%) [25]. The carrier density variation by Ti atom doping is of the order of ~$10^{12}$ cm$^{-2}$, smaller than K atom doping which can reach ~$10^{13}$ cm$^{-2}$.

The Ti-induced charge transfer ($\Delta\rho_{Ti}$) can be calculated by the formula $\Delta\rho_{Ti,exp} = \Delta n/\Delta N_{Ti}$, where $\Delta n$ is the carrier density and $\Delta N_{Ti}$ is the Ti adatom concentration.

Using the charge carrier density increment $\Delta n = 2.5 \times 10^{12}$ $cm^{-2}$ from 0 to 1/235 ML Ti and an adatom concentration $\Delta N_{Ti} = (5.6 \pm 0.6) \times 10^{12}$ $cm^{-2}$, the resulting effective charge transfer is $\Delta\rho_{Ti,exp} = (0.45 \pm 0.05)$ electrons per Ti adatom to epitaxial graphene. This value is twice larger than that previously reported for transition metal adatoms (Ti 0.08 to 0.17, Fe 0.017 to 0.040, and Pt 0.014 to 0.021 electrons) [26].

The orbital-resolved density of states and the bonding electron distribution are calculated by adopting a first-principle approach. The results are shown in Fig. 3. The first-principle calculations are performed using the density functional theory implemented by the Vienna *Ab initio* Simulation Package (VASP). The supercell uses a 4×4 supercell sampling with a localized basis containing 32 carbon atoms (Fig. 3(a)). The optimized geometrical structure is used to calculate the local charge distribution and the orbital hybridization along the slice of **a**+**b**. In this notation, the axes of the a and b directions are placed parallel to and the c direction perpendicular to the graphene plane.

Fig. 3(b) shows the partial density of states (DOS) for relevant atomic orbitals of one Ti adatom on graphene. The Ti orbitals $d_{z^2}$, $d_{xy}$ and $d_{x^2-y^2}$ contribute to the density of states. The electron probability distributions are illustrated in the insets. In particular, the $d_{z^2}$ orbital contributes significantly with states near the Fermi energy. We also note that the $d_{z^2}$ orbital has a similar structure as the vertically distributed C $p_z$ orbital, which enables the adatom to adsorb on the hollow site and not to break the six-fold symmetry of the atomic structure. This results in the $d_{z^2}$ orbital strongly interacting with graphene via lateral orbital coupling. This interaction with graphene is associated with the lowest migration and adsorption energy, which makes it the most stable structure at the hollow site [27]. There are also $d_{xy}$ and $d_{x^2-y^2}$ orbitals which cause weak interaction with graphene.

The bonding electron distribution (BED) shown in Fig. 3(c) is calculated to comprehend how the $d_{z^2}$ orbital interacts with the $p_z$ orbitals. The bonding electron distribution $\Delta\rho$ is computed using the relation

$$\Delta\rho = \rho\,(\text{Ti+Gr}) - \rho\,(\text{Gr}) - \rho\,(\text{Ti}) \qquad (1)$$

where $\rho(\text{Ti+Gr})$ is the local electron distribution for a Ti adatom adsorbed on graphene, while $\rho\,(\text{Gr})$ and $\rho\,(\text{Ti})$ are two independent electron distributions for bare graphene and Ti atoms. The BED accounts for the charge redistribution due to the Ti adatom on graphene. The non-interacting $d_{z^2}$ and $p_z$ orbitals at the Ti adatom and the C atom, respectively, are illustrated by the dashed lines in Fig. 3(c). Here, $3d_{z^2}$ is the primary orbital hybridized with the $2p_z$ orbital out of the supercell plane. It contributes conduction electrons to nearest-neighbor C atoms. Close inspection of orbital hybridization reveals that there are regions of charge accumulation with a positive value (red and yellow solid contour bars) and of depletion with a negative value (blue solid contour bars). The charge significantly transfers from the Ti adatom to the C atoms. In addition, the $3d_{xy}$ and $3d_{x^2-y^2}$ orbitals contribute with minor electron transfer as well. This transferred charge results in the carrier density increase observed by ARPES.

To quantify the charge accumulation in graphene, the charge transfer curve is calculated as a function of Ti coverage. As shown in Fig. 3(d), the transferred charge gradually increases to a maximum value of electron donation as the Ti coverage decreases. At coverages below 0.1 ML, the electron donation reaches $\Delta\rho_{Ti} = 1.1$~$1.2$ per Ti adatom as the maximum charge transfer value. Comparing the experimental $\Delta\rho_{Ti}$ value of about 0.45 from ARPES spectra, this higher theoretically predicted $\Delta\rho_{Ti}$ value is understandable considering three-dimensional (3D)-islanding by a clustering process of Ti adatoms on graphene at room temperature. The Ti adatom clustering process is inevitable at room temperature due to Ti has the large diffusion length [20]. Nevertheless, the DFT calculation result emphasizes the 3D Ti clusters indeed weaken the Ti-C orbital hybridization and significantly reduce the efficiency of Ti-adsorbate induced graphene n-type doping.

In addition, the MDCs linewidths can be used to understand many-body effects in pristine and Ti-doped graphene systems. The MDCs linewidths are extracted from ARPES spectra and displayed as a function of energy by the data points in Fig. 4. The many-body interaction can be understood as follows: the photoexcited hole in the valence band decays through electron-electron and electron-phonon interaction, the former of which includes plasmons and electron-hole pair creation [28,29]. The total momentum linewidth can be defined as $\Delta k = \Delta k_{pl} + \Delta k_h + \Delta k_{ph}$, where $\Delta k_{pl}$ corresponds to the plasmon-, $\Delta k_h$ to the hole-, and $\Delta k_{ph}$ to the phonon-interaction. The MDCs linewidths curve for pristine graphene shows a pronounced peak which is similar to the peak observed in K-doped graphene [25]. This peak is attributed to the electron-plasmon interaction, which is kinematically allowed in the electronic structure through interband transitions [29,30,31,32]. Furthermore, there is a significant peak shift and an enhancement observed for Ti-doped graphene. Thus, as the coverage of Ti adatoms increases, the electron-plasmon interaction shifts towards higher deexcitation energies, which is associated with the larger interband transition energy required for plasmon excitation in the Ti-doped system.

The electron-plasmon model also supports the Dirac cone renormalization and the increased carrier density in Ti-doped graphene. In the long-wavelength limit, the Dirac plasmon energy can be expressed as $\omega_{pl} = \sqrt{\frac{8qE_F \sigma_{uni}}{\hbar \varepsilon}}$ [33,34,35,36], where $\sigma_{uni}$ is the Drude weight, $\varepsilon$ the dielectric constant of the substrate, and $q$ the transferred momentum in the electron-electron interaction. For our doped epitaxial graphene, the transferred momentum is $q = 0.1\ k_F \approx 4.04 \times 10^5\ \text{cm}^{-1}$, the Fermi energy $E_F$ = 360 meV, and the dielectric constant $\varepsilon_{SiC}$ = *7.26*. We therefore estimate the Dirac plasmon energy to be around 177 meV, which is consistent with the fact that the plasmon excitation occurs in graphene/SiC in the infrared range from 50 to 200 meV [37,38]. The dip in the theoretical MDCs is an artefact of the simplicity of the many-body model, which does not include the interaction between plamons and Fermi-liquid excitations. Nevertheless, as shown in Fig. 4, the MDCs of Ti-doped graphene can be quite well fitted to the electron-plasmon model even without considering Fermi-liquid excitations. More details are provided in the supplementary material.

## 4. Conclusions

In conclusion, we have presented ARPES data showing that the deposition of small amounts of Ti on graphene leads to an appreciable band structure modification and velocity renormalization. DFT calculations indicate that the renormalization and doping effects are caused by the lateral coupling of graphene $2p_z$ and Ti $3d_{z^2}$ orbitals. Deposition of Ti adatoms is a simple but highly efficient way to alter the electronic structure of graphene at room-temperature, which enables further development of the application of Dirac cone renormalization.

**Acknowledgements**

We thank to Dr. Yeu-Kuang Hwu, Dr. Chia-Hsin Wang and Dr. Yaw-Wen Yang, the former one from Academia Sinica which provided the mu-metal chamber for ARPES/QCM measurement, and the latter two from NSRRC for their skillful assistance during the experiments at NSRRC "BL 24 A1". The authors would also like to thank Prof. Ming-Fa Lin of NCKU. This work was partially supported by a grant from the Ministry of Science and Technology in Taiwan and in part by the European Union Seventh Framework Programme under Grant Agreement No. 696656 Graphene Flagship Core 1. SH acknowledges financial support from the CNR in the framework of the agreement on scientific collaborations between CNR and CNRS (France), NRF (Korea), and RFBR (Russia); and from the Italian Ministry of Foreign Affairs (Ministero degli Affari Esteri, Direzione Generale per la Promozione del Sistema Paese) in the framework of the agreement on scientific collaborations with Poland.

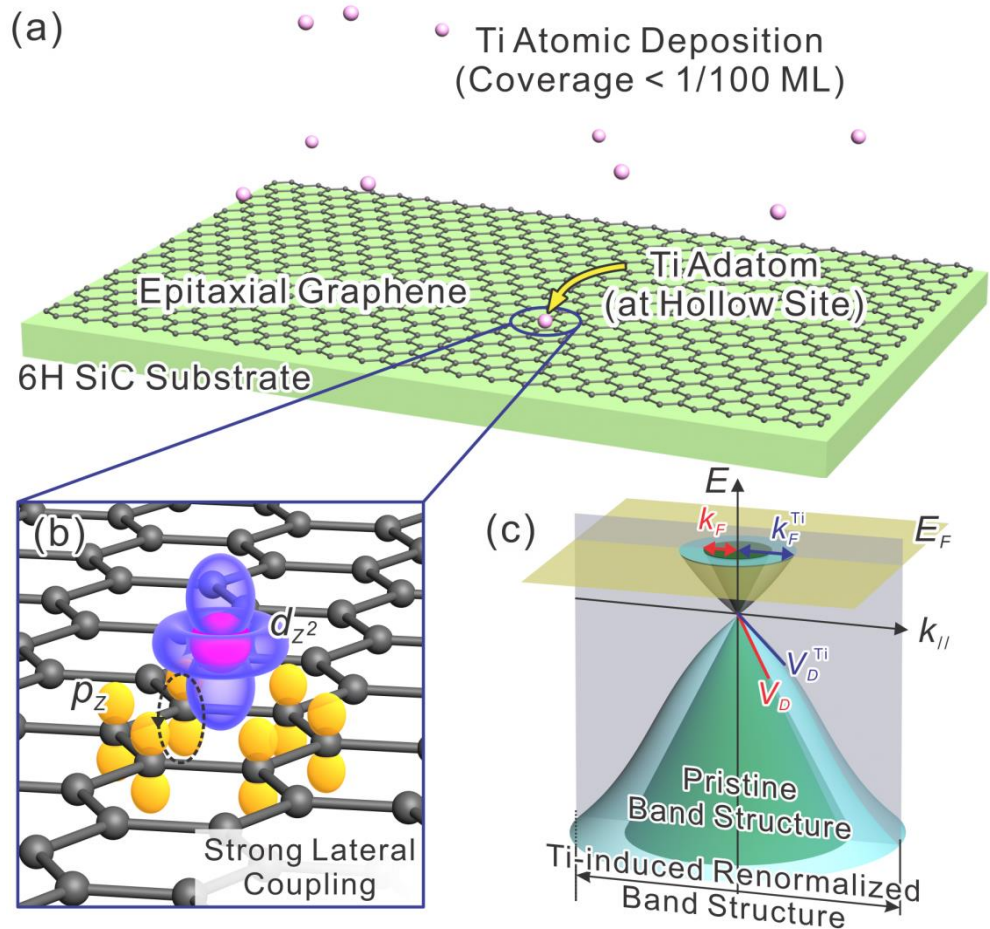

**Figure 1 (color).** Schematics of Ti adatoms deposited onto graphene, the lateral coupling that occurs between the $2p_z$ orbital of graphene the $3d_{z^2}$ orbital of Ti, and the resulting band structure renormalization. (a) Atomic deposition of Ti onto single-layer epitaxial graphene. (b) Ti-$3d$ and C-$2p_z$ orbital hybridization. (c) Electronic structure illustration of the band renormalization on Ti-decorated graphene. The electron-pocket is strongly influenced after Ti adsorption.

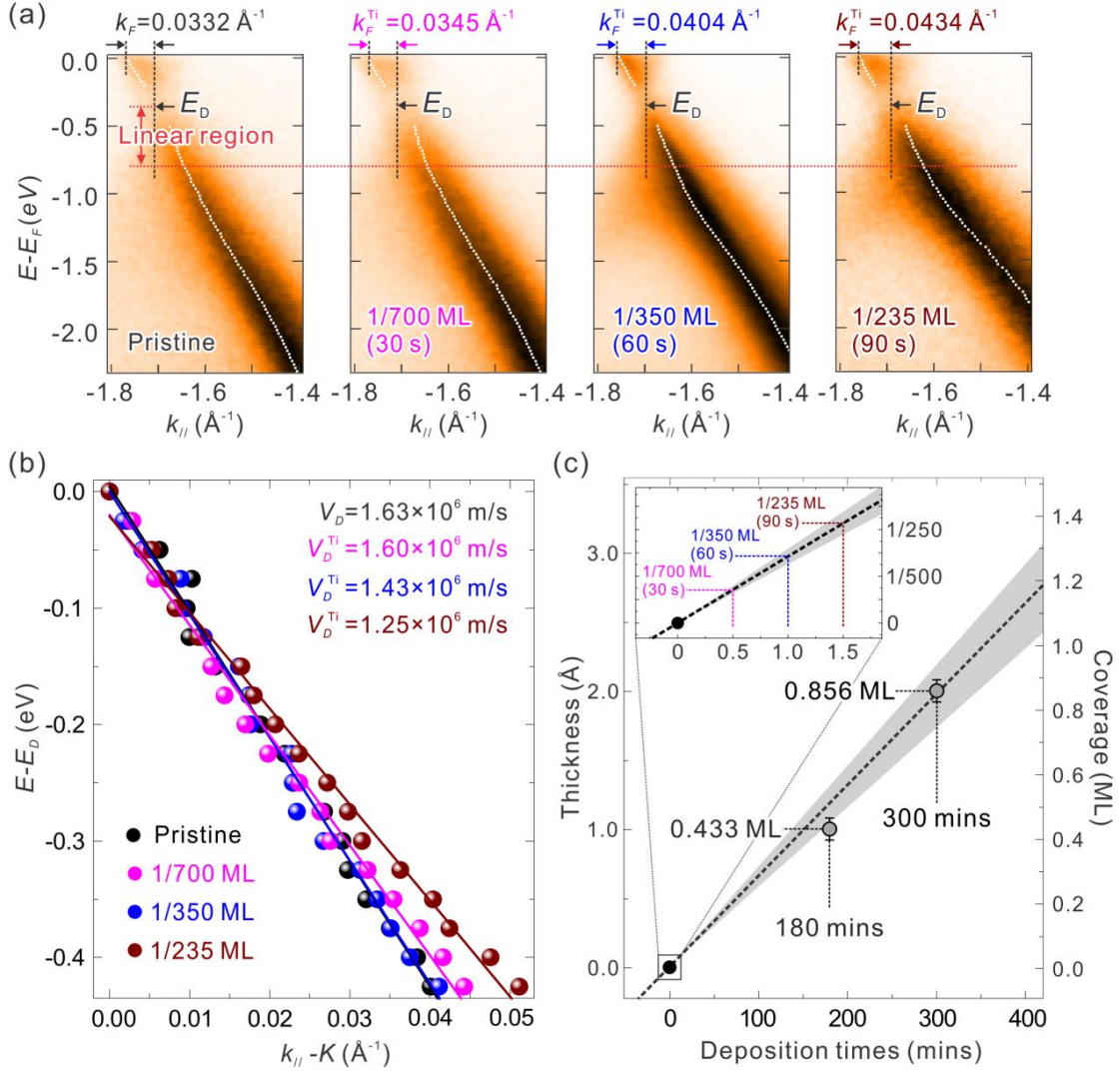

**Figure 2 (color).** ARPES measurements on graphene with Ti adatoms using an incoming SR photon energy of 52 eV at room temperature. (a) Band structure of Ti-doped graphene for 0, 1/700, 1/350, and 1/235 ML coverage, taken along the **K→Γ** direction and in the vicinity of the K-point. The energy dispersion is fitted by the MDCs (white dashed lines). The black arrow marks the binding energy position of the Dirac point. The acquired Fermi momentum $k_F$ values are given. (b) Dirac velocity in the low-lying valence band depends on Ti coverage. (c) The Ti coverage calibration obtained by QCM. Deposition conditions and corresponding Ti coverages are listed. The black dashed line shows a least squares fit to the data.

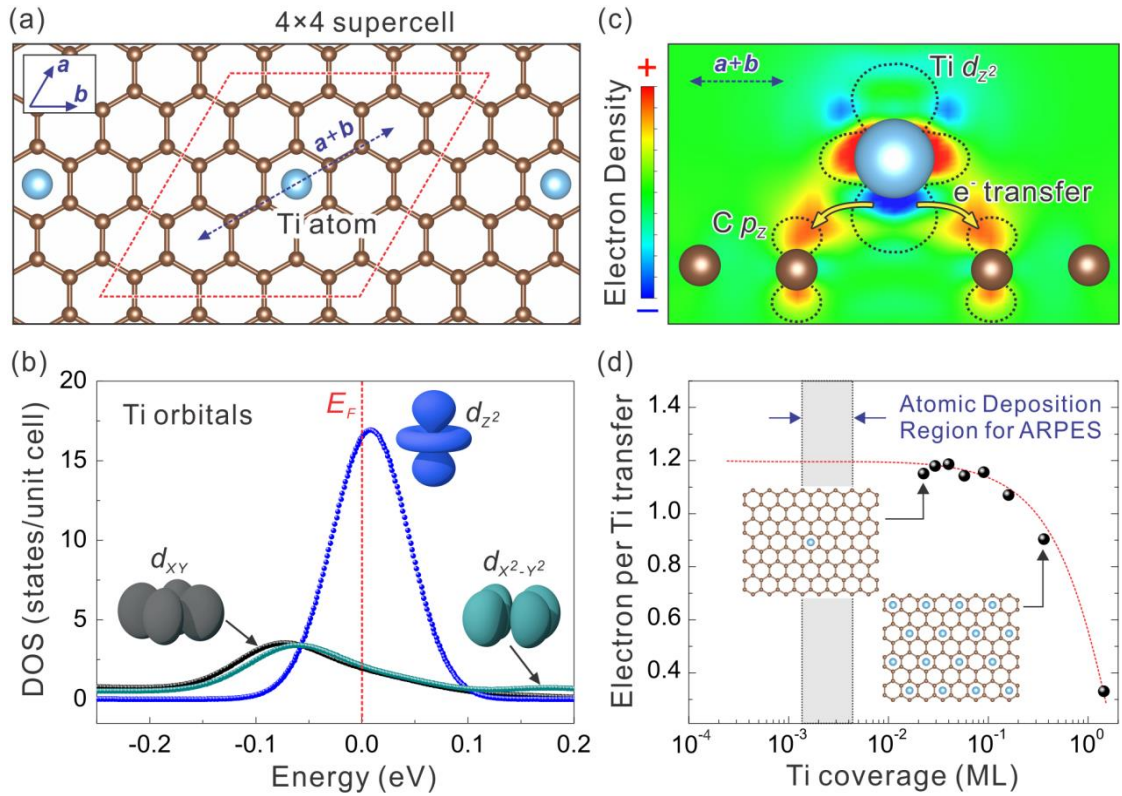

**Figure 3 (color).** First-principles DFT calculations: orbital contribution, bonding electron distribution, and charge transfer from Ti adatom to epitaxial graphene. (a) The supercell of carbon atoms is taken as a 4×4 supercell with a localized basis. The geometrical structure is used to determine the local charge distribution along the slice **a**+**b**. (b) Atomic orbitals of $3d_{z^2}$, $3d_{xy}$ and $3d_{x^2-y^2}$ which contribute to the density of states (DOS). Electron probability distributions are shown as insets. The $3d_{z^2}$ orbital is the primary orbital which hybridizes with the $2p_z$ orbital normal to the supercell plane. (c) The bonding electron distribution (BED) for Ti adatoms on epitaxial graphene. The non-interacting orbitals distribution of $3d_{z^2}$ and $2p_z$ is depicted as a dashed line. (d) Charge transfer per Ti atom as a function of Ti adatom coverage. Insets illustrate the different amounts of Ti adatom concentration.

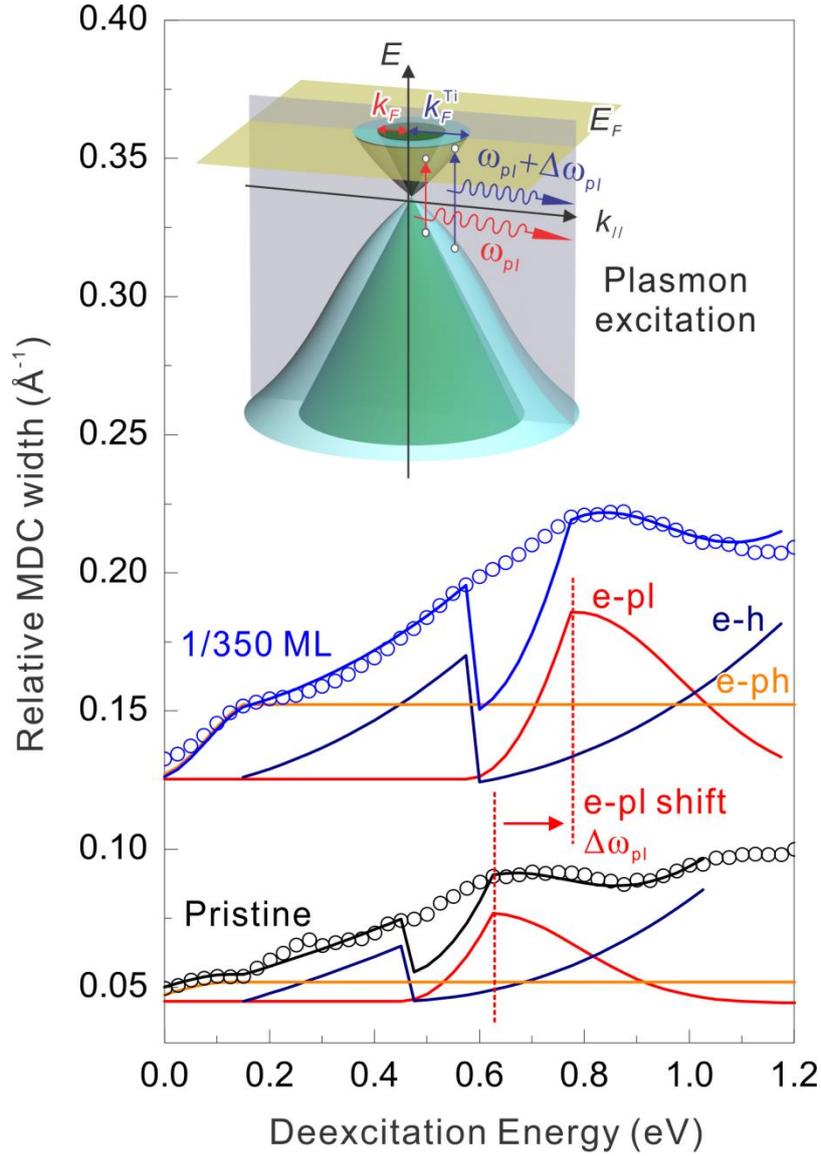

**Figure 4 (color).** MDCs linewidths used to extract the many-body interactions for pristine and 1/350 ML Ti-doped epitaxial graphene. The trace corresponding to the doped graphene is shifted upwards by 0.13 Å$^{-1}$. MDCs line-shape analysis is carried out by electron-phonon, electron-hole, and electron-plasmon interactions. The electron-plasmon peak shift is associated with the renormalized electronic structure shown in the inset.